\documentclass[twocolumn,letterpaper,aps,prl,superscriptaddress,showpacs,floatfix]{revtex4-1}
\usepackage{graphicx}   % Include figure filed

\newcommand{\mean}[1]{\left\langle #1 \right\rangle}

\begin{document}

%%%%%%%%%%%%%%%%%%%% Title and Authors %%%%%%%%%%%%%%%%%%%%%%%%%%%%%%%%%%%%%%

\title{Measurements of Higher-Order Flow Harmonics in Au$+$Au Collisions \\
		at $\sqrt{s_{NN}}=200$~GeV }

\newcommand{\abilene}{Abilene Christian University, Abilene, Texas 79699, USA}
\newcommand{\banaras}{Department of Physics, Banaras Hindu University, Varanasi 221005, India}
\newcommand{\barc}{Bhabha Atomic Research Centre, Bombay 400 085, India}
\newcommand{\bnlcoll}{Collider-Accelerator Department, Brookhaven National Laboratory, Upton, New York 11973-5000, USA}
\newcommand{\bnlphys}{Physics Department, Brookhaven National Laboratory, Upton, New York 11973-5000, USA}
\newcommand{\caucr}{University of California - Riverside, Riverside, California 92521, USA}
\newcommand{\charlesczech}{Charles University, Ovocn\'{y} trh 5, Praha 1, 116 36, Prague, Czech Republic}
\newcommand{\chonbuk}{Chonbuk National University, Jeonju, 561-756, Korea}
\newcommand{\ciae}{Science and Technology on Nuclear Data Laboratory, China Institute of Atomic Energy, Beijing 102413, P.~R.~China}
\newcommand{\cns}{Center for Nuclear Study, Graduate School of Science, University of Tokyo, 7-3-1 Hongo, Bunkyo, Tokyo 113-0033, Japan}
\newcommand{\colorado}{University of Colorado, Boulder, Colorado 80309, USA}
\newcommand{\columbia}{Columbia University, New York, New York 10027 and Nevis Laboratories, Irvington, New York 10533, USA}
\newcommand{\czechtech}{Czech Technical University, Zikova 4, 166 36 Prague 6, Czech Republic}
\newcommand{\dapnia}{Dapnia, CEA Saclay, F-91191, Gif-sur-Yvette, France}
\newcommand{\debrecen}{Debrecen University, H-4010 Debrecen, Egyetem t{\'e}r 1, Hungary}
\newcommand{\elte}{ELTE, E{\"o}tv{\"o}s Lor{\'a}nd University, H - 1117 Budapest, P{\'a}zm{\'a}ny P. s. 1/A, Hungary}
\newcommand{\ewha}{Ewha Womans University, Seoul 120-750, Korea}
\newcommand{\fit}{Florida Institute of Technology, Melbourne, Florida 32901, USA}
\newcommand{\fsu}{Florida State University, Tallahassee, Florida 32306, USA}
\newcommand{\gsu}{Georgia State University, Atlanta, Georgia 30303, USA}
\newcommand{\hiroshima}{Hiroshima University, Kagamiyama, Higashi-Hiroshima 739-8526, Japan}
\newcommand{\ihepprot}{IHEP Protvino, State Research Center of Russian Federation, Institute for High Energy Physics, Protvino, 142281, Russia}
\newcommand{\illuiuc}{University of Illinois at Urbana-Champaign, Urbana, Illinois 61801, USA}
\newcommand{\inrras}{Institute for Nuclear Research of the Russian Academy of Sciences, prospekt 60-letiya Oktyabrya 7a, Moscow 117312, Russia}
\newcommand{\instpasczech}{Institute of Physics, Academy of Sciences of the Czech Republic, Na Slovance 2, 182 21 Prague 8, Czech Republic}
\newcommand{\isu}{Iowa State University, Ames, Iowa 50011, USA}
\newcommand{\jinrdubna}{Joint Institute for Nuclear Research, 141980 Dubna, Moscow Region, Russia}
\newcommand{\jyvaskyla}{Helsinki Institute of Physics and University of Jyv{\"a}skyl{\"a}, P.O.Box 35, FI-40014 Jyv{\"a}skyl{\"a}, Finland}
\newcommand{\kek}{KEK, High Energy Accelerator Research Organization, Tsukuba, Ibaraki 305-0801, Japan}
\newcommand{\kfki}{KFKI Research Institute for Particle and Nuclear Physics of the Hungarian Academy of Sciences (MTA KFKI RMKI), H-1525 Budapest 114, POBox 49, Budapest, Hungary}
\newcommand{\korea}{Korea University, Seoul, 136-701, Korea}
\newcommand{\kurchatov}{Russian Research Center ``Kurchatov Institute", Moscow, 123098 Russia}
\newcommand{\kyoto}{Kyoto University, Kyoto 606-8502, Japan}
\newcommand{\labllr}{Laboratoire Leprince-Ringuet, Ecole Polytechnique, CNRS-IN2P3, Route de Saclay, F-91128, Palaiseau, France}
\newcommand{\lawllnl}{Lawrence Livermore National Laboratory, Livermore, California 94550, USA}
\newcommand{\losalamos}{Los Alamos National Laboratory, Los Alamos, New Mexico 87545, USA}
\newcommand{\lpc}{LPC, Universit{\'e} Blaise Pascal, CNRS-IN2P3, Clermont-Fd, 63177 Aubiere Cedex, France}
\newcommand{\lund}{Department of Physics, Lund University, Box 118, SE-221 00 Lund, Sweden}
\newcommand{\maryland}{University of Maryland, College Park, Maryland 20742, USA}
\newcommand{\mass}{Department of Physics, University of Massachusetts, Amherst, Massachusetts 01003-9337, USA }
\newcommand{\muenster}{Institut fur Kernphysik, University of Muenster, D-48149 Muenster, Germany}
\newcommand{\muhlenberg}{Muhlenberg College, Allentown, Pennsylvania 18104-5586, USA}
\newcommand{\myongji}{Myongji University, Yongin, Kyonggido 449-728, Korea}
\newcommand{\nagasaki}{Nagasaki Institute of Applied Science, Nagasaki-shi, Nagasaki 851-0193, Japan}
\newcommand{\newmex}{University of New Mexico, Albuquerque, New Mexico 87131, USA }
\newcommand{\nmsu}{New Mexico State University, Las Cruces, New Mexico 88003, USA}
\newcommand{\ornl}{Oak Ridge National Laboratory, Oak Ridge, Tennessee 37831, USA}
\newcommand{\orsay}{IPN-Orsay, Universite Paris Sud, CNRS-IN2P3, BP1, F-91406, Orsay, France}
\newcommand{\peking}{Peking University, Beijing 100871, P.~R.~China}
\newcommand{\pnpi}{PNPI, Petersburg Nuclear Physics Institute, Gatchina, Leningrad region, 188300, Russia}
\newcommand{\riken}{RIKEN Nishina Center for Accelerator-Based Science, Wako, Saitama 351-0198, Japan}
\newcommand{\rikjrbrc}{RIKEN BNL Research Center, Brookhaven National Laboratory, Upton, New York 11973-5000, USA}
\newcommand{\rikkyo}{Physics Department, Rikkyo University, 3-34-1 Nishi-Ikebukuro, Toshima, Tokyo 171-8501, Japan}
\newcommand{\saispbstu}{Saint Petersburg State Polytechnic University, St. Petersburg, 195251 Russia}
\newcommand{\saopaulo}{Universidade de S{\~a}o Paulo, Instituto de F\'{\i}sica, Caixa Postal 66318, S{\~a}o Paulo CEP05315-970, Brazil}
\newcommand{\seoulnat}{Seoul National University, Seoul, Korea}
\newcommand{\stonybrkc}{Chemistry Department, Stony Brook University, SUNY, Stony Brook, New York 11794-3400, USA}
\newcommand{\stonycrkp}{Department of Physics and Astronomy, Stony Brook University, SUNY, Stony Brook, New York 11794-3400, USA}
\newcommand{\tenn}{University of Tennessee, Knoxville, Tennessee 37996, USA}
\newcommand{\titech}{Department of Physics, Tokyo Institute of Technology, Oh-okayama, Meguro, Tokyo 152-8551, Japan}
\newcommand{\tsukuba}{Institute of Physics, University of Tsukuba, Tsukuba, Ibaraki 305, Japan}
\newcommand{\vandy}{Vanderbilt University, Nashville, Tennessee 37235, USA}
\newcommand{\waseda}{Waseda University, Advanced Research Institute for Science and Engineering, 17 Kikui-cho, Shinjuku-ku, Tokyo 162-0044, Japan}
\newcommand{\weizmann}{Weizmann Institute, Rehovot 76100, Israel}
\newcommand{\yonsei}{Yonsei University, IPAP, Seoul 120-749, Korea}
\affiliation{\abilene}
\affiliation{\banaras}
\affiliation{\barc}
\affiliation{\bnlcoll}
\affiliation{\bnlphys}
\affiliation{\caucr}
\affiliation{\charlesczech}
\affiliation{\chonbuk}
\affiliation{\ciae}
\affiliation{\cns}
\affiliation{\colorado}
\affiliation{\columbia}
\affiliation{\czechtech}
\affiliation{\dapnia}
\affiliation{\debrecen}
\affiliation{\elte}
\affiliation{\ewha}
\affiliation{\fit}
\affiliation{\fsu}
\affiliation{\gsu}
\affiliation{\hiroshima}
\affiliation{\ihepprot}
\affiliation{\illuiuc}
\affiliation{\inrras}
\affiliation{\instpasczech}
\affiliation{\isu}
\affiliation{\jinrdubna}
\affiliation{\jyvaskyla}
\affiliation{\kek}
\affiliation{\kfki}
\affiliation{\korea}
\affiliation{\kurchatov}
\affiliation{\kyoto}
\affiliation{\labllr}
\affiliation{\lawllnl}
\affiliation{\losalamos}
\affiliation{\lpc}
\affiliation{\lund}
\affiliation{\maryland}
\affiliation{\mass}
\affiliation{\muenster}
\affiliation{\muhlenberg}
\affiliation{\myongji}
\affiliation{\nagasaki}
\affiliation{\newmex}
\affiliation{\nmsu}
\affiliation{\ornl}
\affiliation{\orsay}
\affiliation{\peking}
\affiliation{\pnpi}
\affiliation{\riken}
\affiliation{\rikjrbrc}
\affiliation{\rikkyo}
\affiliation{\saispbstu}
\affiliation{\saopaulo}
\affiliation{\seoulnat}
\affiliation{\stonybrkc}
\affiliation{\stonycrkp}
\affiliation{\tenn}
\affiliation{\titech}
\affiliation{\tsukuba}
\affiliation{\vandy}
\affiliation{\waseda}
\affiliation{\weizmann}
\affiliation{\yonsei}
\author{A.~Adare} \affiliation{\colorado}
\author{S.~Afanasiev} \affiliation{\jinrdubna}
\author{C.~Aidala} \affiliation{\mass}
\author{N.N.~Ajitanand} \affiliation{\stonybrkc}
\author{Y.~Akiba} \affiliation{\riken} \affiliation{\rikjrbrc}
\author{H.~Al-Bataineh} \affiliation{\nmsu}
\author{J.~Alexander} \affiliation{\stonybrkc}
\author{K.~Aoki} \affiliation{\kyoto} \affiliation{\riken}
\author{Y.~Aramaki} \affiliation{\cns}
\author{E.T.~Atomssa} \affiliation{\labllr}
% ...
\collaboration{PHENIX Collaboration} \noaffiliation

\date{\today}

%%%%%%%%%%%%%%%%%%%%%%%%%%%%%% Abstract %%%%%%%%%%%%%%%%%%%%%%%%%%%%%%%%%%%%%%

\begin{abstract}

Flow coefficients $v_n$ for $n$ =~2,~3,~4, characterizing the
anisotropic collective flow in Au+Au collisions at
$\sqrt{s_{NN}}~=~200$ GeV, are measured relative to event planes
$\Psi_{n}$, determined at large rapidity.   We report $v_n$ as a
function of transverse momentum and collision centrality, and study
the correlations among the event planes of different order $n$.  The
$v_n$ are well described by hydrodynamic models which employ a
Glauber Monte Carlo initial state geometry with fluctuations,
providing additional constraining power on the interplay between
initial conditions and the effects of viscosity as the system
evolves.  This new constraint can serve to improve the precision of 
the extracted shear viscosity to entropy density ratio $\eta/s$.

\end{abstract}


\maketitle

%%%%%%%%%%%%%%%%%%%%%%%%%%%%% Introduction %%%%%%%%%%%%%%%%%%%%%%%%%%%%%%%

The production of particles in heavy ion collisions at the
Relativistic Heavy Ion Collider is anisotropic in directions
transverse to the beam.  For low momentum particles ($p_T \alt 3$
GeV/$c$), this anisotropy is understood to result from hydrodynamically
driven flow of the quark-gluon plasma 
(QGP)~\cite{Ollitrault:1992bk,Heinz:2001xi,Shuryak:2008eq,Adcox:2004mh,Adams:2005dq}.
The strength of the flow is measured as Fourier coefficients $v_{n} =
\mean{e^{i\,n(\phi - \Psi_{\rm RP})}}, n=2,4,..$ where $\phi$ is the 
azimuthal angle of an emitted particle around the $z$ axis defined by the 
beam; $\Psi_{\rm RP}$ is the azimuth of the 
reaction plane defined by the beam direction and the impact vector between 
the colliding nuclei. The brackets denote averaging over particles and 
events.  The reaction plane is not measurable directly a priori, so the 
Fourier coefficients are determined with respect to the estimated 
participant event planes~\cite{Ollitrault:1992bk}.  Earlier measurements have
focused on the even-order anisotropies $v_2$ 
and $v_4$, evaluated with respect to an event plane $\Psi_2$, determined 
from the $n=2$ correlation.

The $v_2$($v_4$) values obtained this way for a broad range of
$p_{T}$ and centrality  have been used to extract the specific
viscosity $\eta/s$ (the ratio of shear viscosity $\eta$ to entropy density
$s$) of the hot and dense nuclear matter via hydrodynamic model
comparisons~\cite{Romatschke:2007mq,Hirano:2009ah,Song:2010mg,Adare:2010ux,Gombeaud:2009ye}.  
These model comparisons, which incorporate the dynamic evolution of an
early-stage strongly-coupled QGP, together with a late-stage hadronic
gas, show an ambiguity for very different values of  $4\pi\eta/s \simeq 2$ and
$4\pi\eta/s \simeq 1$, the latter being a conjectured lower bound for the
specific viscosity~\cite{Kovtun:2004de}. Specifically the two values 
correspond to two equally successful parameter sets, each including different
estimates of the initial state anisotropy (parameterized as ``eccentricity''
see below)~\cite{Hirano:2009ah,Song:2010mg,Lacey:2010fe},
which dominate the associated uncertainty in these models.  The 
lower bound value is obtained with a standard Glauber Monte Carlo
(Glauber-MC) model~\cite{Miller:2007ri,Alver:2006wh} of the initial state which
results in smaller initial elliptical eccentricity and thus needs less viscosity
to reproduce the measured final state particle anisotropy. The higher
value $4\pi\eta/s \simeq 2$, corresponds to a larger initial
eccentricity in the color-glass condensate (CGC) inspired 
Monte-Carlo-Kharzeev-Levin-Nardi (MC-KLN) 
model~\cite{Kharzeev:2000ph,Lappi:2006xc,Drescher:2007ax} of the initial state.

Recently, significant attention has been given to the  study of the
influence of initial geometry fluctuations of the initial state
anisotropy~\cite{Alver:2010gr} which are typically quantified by higher-order
generalized ``eccentricities" $\varepsilon_{n}$~\cite{Alver:2010gr,Lacey:2010hw}.  
The goal has been to understand how such fluctuations induce anisotropic 
particle emission, characterized by $v_n$ (for odd and even $n$)
\begin{equation}
\frac{dN}{d\phi} \propto
1 + \sum_{n=1} 2 \, v_n \, \cos(n[\phi-\Psi_n]),  
\label{eq:basic_singles_psi_n}
\end{equation}
where $v_n = \mean{ \cos(n[\phi - \Psi_{n}])}, n=1,2,3,...$
and the $\Psi_n$ are the generalized participant event
planes at all orders for each event.   These recent developments
suggest that measurements of $v_n$, especially for $n$ = 3, can yield
important additional constraints that provide a more precise estimate
of $\frac{\eta}{s}$, as well as resolve the correct eccentricity
model.

Here we present results for differential measurements
following Eq.~\ref{eq:basic_singles_psi_n}, for Au+Au collisions at
$\sqrt{s_{NN}}$=200~GeV.  We first show how the measured event planes
correlate across large rapidity gaps, and then show resulting $v_n$
moments for midrapidity particles relative to those planes.
The results are derived from $\sim 3.0 \times 10^9$  Au+Au events
obtained with the PHENIX detector~\cite{Adcox:2003zm} during the 2007
running period.  Collision centrality (related to impact parameter)
and number of participating nucleons ($N_{\rm part}$) are estimated as 
in~\cite{Adare:2010ux} through comparisons of detected multiplicity in
beam-beam counters (BBC)~\cite{Allen:2003zt} with
a Glauber-MC calculation.  Event planes were determined
using three separate detector systems: the same BBCs,
reaction-plane detectors
(RXN)~\cite{Richardson:2010hm}, and muon-piston calorimeters (MPC).
Each detector system has a North (South) component to measure at
forward (backward) rapidity.  The absolute pseudorapidity ($\eta'$)
coverage for these detectors are $3.1 < \left| \eta'_{_{\rm BBC}}
\right| < 3.9 $, $1.0 < \left| \eta'_{{\rm RXN}} \right| < 2.8$, $3.1
< \left| \eta'_{_{\rm MPC}} \right| < 3.7 $.  The PHENIX drift and
pad chambers~\cite{Adcox:2003zp} were used for charged
particle tracking and momentum reconstruction with azimuthal angle coverage
$\varphi=\pi$ rad in the central region ($|\eta'|\leq 0.35$).

To estimate the event plane $\Psi_n$ in each detector, we generalize
to all orders $n$ our earlier procedure for event plane determination 
(see~\cite{Adare:2010ux} and especially definitions 
in~\cite{Afanasiev:2009wq}). For each event plane
detector we evaluate 
$\tan(n \Phi_{n}) = \sum w_i\sin(n \phi_i) 
/ \sum w_i\cos(n \phi_i)$ 
for the $\Psi_n$ subevent estimator $\Phi_n$,
where the $\phi_{i}$ are the azimuths of elements in that detector
and the weights $w_{i}$ reflect the energy or multiplicity in that
element. Acceptance corrections~\cite{Afanasiev:2009wq} for imperfect
detector efficiency were employed to ensure a flat (azimuthally independent) 
event plane distribution, as required by symmetry considerations.  
In general, the hit distributions sample virtually all momenta.

%%%%%%%%%%%%%%%%%%%%%%%%%%%%%%%%%%%%%%%%%%%%%%%  Fig_1
\begin{figure}[t]
\includegraphics[width=1.0\linewidth]{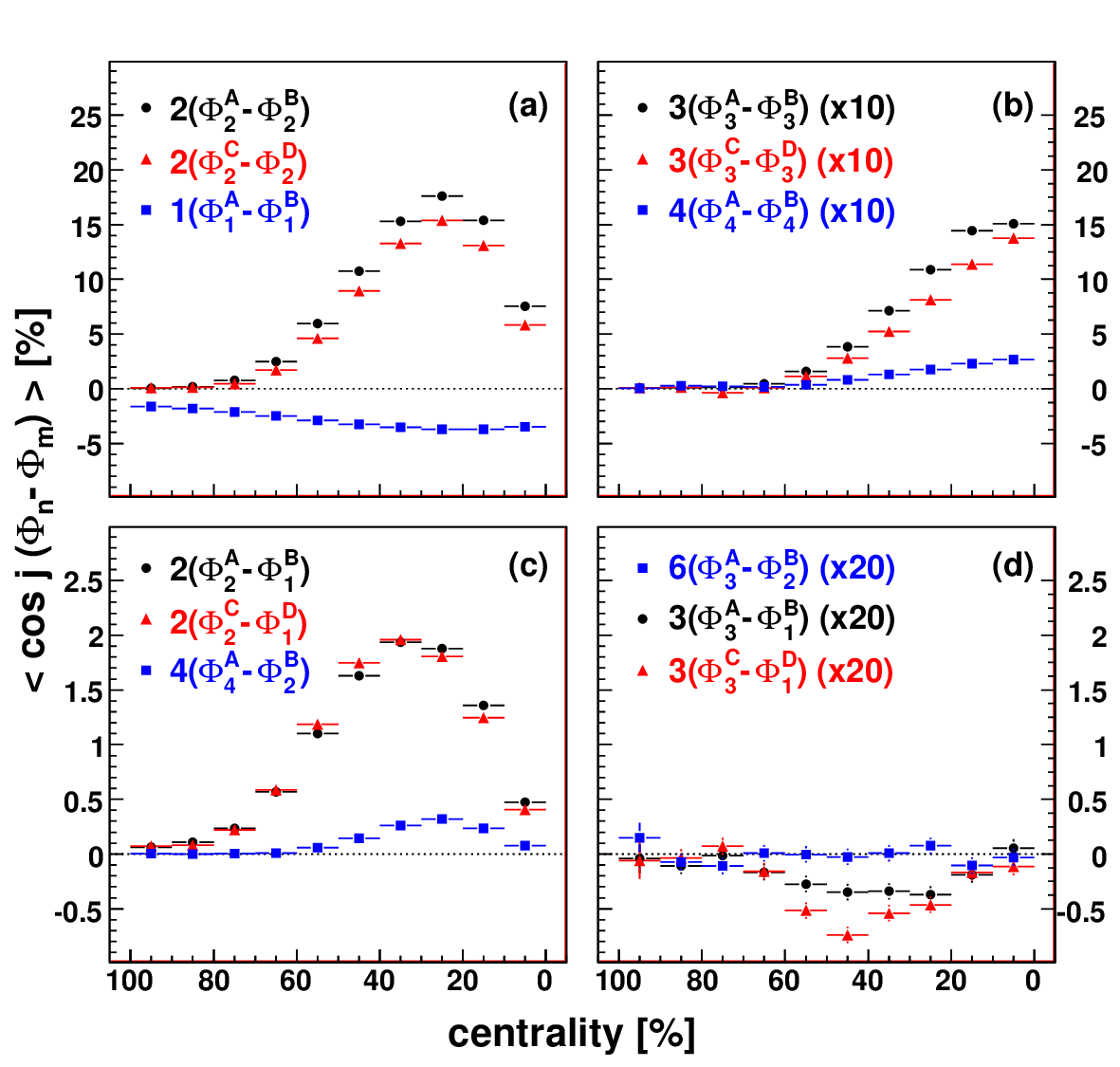}
\caption{(color online). 
Raw correlation strengths 
$\left\langle \cos(j[\Phi_n^{A}-\Phi_m^{B}])\right\rangle$
and $\left\langle \cos(j[\Phi_n^{C}-\Phi_m^{D}])\right\rangle$
of the event planes for various 
detector combinations as a function of the collision centrality, binned
in percentages of the total cross section, where 0 \% corresponds to
impact parameter = 0.  Panels (a) and (b) show
the two sub-events correlations for $m = n$; (c) and (d) show the
two sub-events correlations for $m \neq n$.
The detectors in which the event plane is measured 
are: A: RXN North, B: BBC South, C: MPC North, and D: MPC South.  Data
in (b) and (d) have been scaled by factors of 10 and 20, respectively.  
}
\label{Fig1}
\end{figure}

To measure $v_n$, the azimuth $\phi$ of each particle is correlated
with the $\Psi_n$ via Eq. 1.  The measured $v_{n}\{\Psi_n\} = 
{\left\langle \cos(n[\phi-\Phi_n^{{\rm avg}}])\right\rangle}
       / {{\rm Res}(\Psi_{n})}$, where $\Phi_{n}^{{\rm avg}}$ is
the average of the $\Phi_n$ for North and South subevents and where
the denominator Res($\Psi_n$) represents a resolution factor
described in~\cite{Afanasiev:2009wq}. This factor corrects $v_n$ for
the event-by-event dispersion of the $\Phi_n$. Its magnitude can be
estimated via the two and three sub-events method~\cite{Adare:2010ux}
in which the correlation between $\Phi_n$ from different sub-events
is measured.  The strength of this correlation is generally
quantified as $\left\langle 
\cos(n[\Phi_n^{A}-\Phi_n^{B}])\right\rangle$ for sub-events $A,B$, 
which measures the cosine of the dispersion of the $\Phi_n$ estimator 
with respect to the true $\Psi_n$.

Figure~\ref{Fig1} shows the centrality dependence of this correlation 
strength $\left\langle \cos(j[\Phi_n^{A}-\Phi_m^{B}])\right\rangle$ 
for sub-event combinations ($A,B$)
involving different event-plane detectors with $\Delta\eta' \sim 5$
and  $\Delta\eta' \sim 7$.  The raw correlations are presented as
measured; however, the magnitudes are specific
to the PHENIX detectors involved.  The systematic uncertainties (not
shown) for these correlations are of similar relative size to those
for $v_{n}\{\Psi_n\}$ discussed below.  The uncertainties are
correlated across centrality and $n$ such that the relative size of
these event plane correlations can be compared. The magnitudes for
the odd parity quantities 
$\left\langle \sin(j[\Phi_n^{A}-\Phi_m^{B}])\right\rangle$, 
which should vanish, are found to be
consistent with zero for all centrality, $j$, and $\Phi$
combinations.
Figure~\ref{Fig1} panels (a) and (b) show 
the two sub-events correlations for $m = n$; (c) and (d) show the 
two sub-events correlations for $m \neq n$.   
The negative correlation indicated in (a) for $n=1$ is due
to the well known antisymmetric pseudorapidity dependence (sign change
about midrapidity) of sidewards 
flow $v_1$, as well as momentum conservation~\cite{Heinz:2001xi}.  
Positive sub-event correlations are indicated
in (a) and (b) for $\Psi_{2,3,4}$, with sizable magnitudes
for $\Psi_{2,3}$ and much smaller values for $\Psi_4$.

The sub-event correlations 
$\left\langle \cos(j[\Phi_n^{A}-\Phi_m^{B}])\right\rangle$ 
for $n \neq m$ are also of interest.
Figure 1(c) confirms the expected correlation between $\Psi_1$ and
$\Psi_2$ (due to sidewards flow), as well as that between $\Psi_2$
and $\Psi_4$~\cite{Afanasiev:2009wq}. By contrast, Fig. 1(d) shows
that there is no significant correlation observed between $\Psi_2$
and $\Psi_3$, a result which is independent of the detectors used.
The order $j = 6$ is chosen to account for the
$n$-multiplet of directions ($2\pi/n$) of $\Psi_2$ and $\Psi_3$. The
absence of this correlation suggests that the fluctuations for
$\Psi_3$ about $\Psi_2$ are substantial. This is well reproduced by
Glauber modeling~\cite{Nagle:2010zk,Lacey:2010av} and therefore
supports an initial state fluctuation origin of $\Psi_3$ and $v_3$. A
small correlation between $\Psi_3$ and $\Psi_1$ is indicated in 
Fig.~1(d). While such a correlation seems to be at odds with the absence
of a $\Psi_2 -\Psi_3$ correlation [Fig. 1(d)], we note that $\Psi_1
- \Psi_3$ correlations need not contribute to a residual contribution
to $\Psi_2 - \Psi_3$ correlations through $\Psi_1$. That is, $\Psi_1$
could correlate with $\Psi_3$ and $\Psi_2$ in exclusive event
classes.  Correlations involving the PHENIX zero-degree calorimeter, 
which measures the $n$ = 1 spectator neutron event 
plane~\cite{Afanasiev:2009wq} at $|\eta'| > 6.5$ indicate that this
correlation has some degree of $\eta'$-antisymmetry.  We defer further
investigation of these correlation subtleties to future work.

\begin{figure*}[t]
\includegraphics[width=0.99\linewidth]{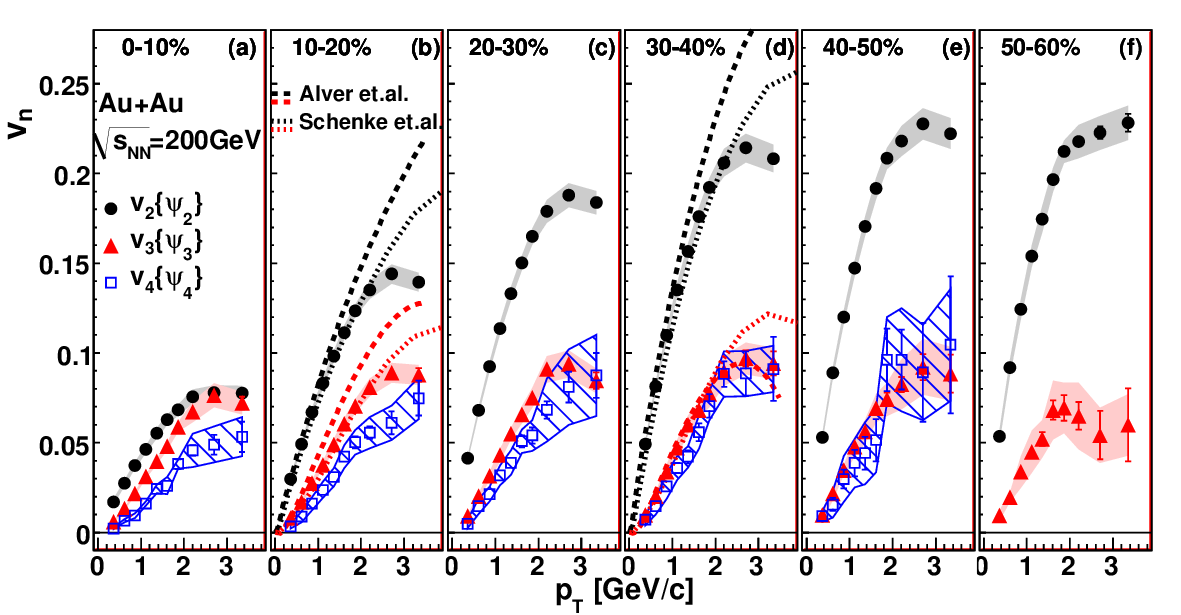}
\caption{(color online). 
$v_{n}\{\Psi_n\}$ vs $p_T$ measured via the reaction plane method 
for different centrality bins; 0\%--10\% are the most central 
collisions.  Shaded (gray and pink) and hatched (blue) areas around the 
data points indicate sizes of systematic uncertainties.
The curves in panels (b) and (d) are predictions 
for $v_{2}\{\Psi_2\}$ and $v_{3}\{\Psi_3\}$ from two 
hydrodynamic models, both using 
Glauber initial conditions and $4\pi\eta/s = 1$, 
Alver {\it et al.}~\cite{Alver:2010dn} and 
Schenke {\it et al.}~\cite{Schenke:2010rr}.
}
\label{Fig2}
\end{figure*}

Figure~\ref{Fig2} shows results for the midrapidity $v_{n}\{\Psi_n\}$ 
for tracks in the central arms as a function of $p_T$ for
different centralities. RXN-defined event planes, which have the best
resolution, are employed.  The
systematic uncertainties for these measurements were estimated by
detailed comparisons of the results obtained with the RXN, BBC, and
MPC event-plane detectors and subevent selections. They are $\sim
3$\%, $\sim 8$\% and $\sim 20$\% for $v_{2}\{\Psi_2\}$,
$v_{3}\{\Psi_3\}$, and $v_{4}\{\Psi_4\}$, respectively, for
midcentral collisions and increase by a few percent for more central
and peripheral collisions. Through further comparison of the results
obtained with the RXN, BBC, and MPC event plane detectors,
pseudorapidity dependent nonflow contributions that may influence
the magnitude of $v_{n}\{\Psi_n\}$, such as jet correlations, were
shown~\cite{Adare:2010ux} to be much less than all other uncertainties for
$v_{2}\{\Psi_2\}$ and $v_{4}\{\Psi_2\}$.

The $v_{n}\{\Psi_n\}$ values shown in Fig.~\ref{Fig2} increase with
$p_T$ for most of the measured range, and decrease for more central
collisions. The $v_2\{\Psi_{2}\}$ increases as expected from central 
to semi-peripheral collisions, following the expected increase of 
$\varepsilon_n$ with impact parameter 
~\cite{Alver:2010dn,Staig:2010pn,Lacey:2010hw}. The $v_{3}\{\Psi_3\}$ 
and, albeit with less statistical significance, also the $v_4\{\Psi_4\}$
appear to be much less centrality dependent, with $v_{3}$ values comparable 
to $v_{2}\{\Psi_2\}$ in the most central events.  
This behavior is consistent with  Glauber calculations of the average
fluctuations of the generalized ``triangular" eccentricity
$\varepsilon_3$~\cite{Nagle:2010zk,Lacey:2010av}.  The Fig.~\ref{Fig2}
panels (b) and (d) show comparisons of $v_{2}\{\Psi_2\}$ and
$v_{3}\{\Psi_3\}$ to results from hydrodynamic calculations.  The
$p_T$ and centrality trends for both $v_{2}\{\Psi_2\}$ and
$v_{3}\{\Psi_3\}$ are in good agreement with the hydrodynamic models
shown, especially at $p_T$ below $\approx 1$~GeV/$c$.

%%%%%%%%%%%%%%%%%%%%%%%%%%%%%%%%%  Fig_2

\begin{figure}[t]
\includegraphics[width=1.0\linewidth]{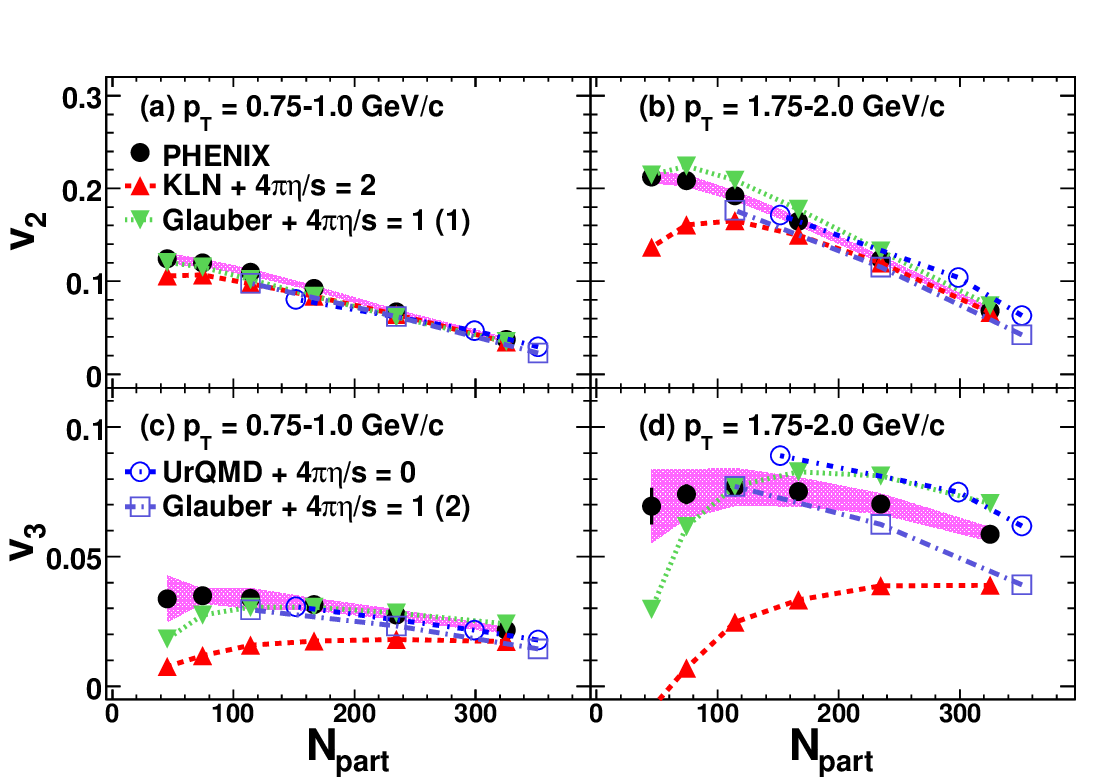}
\caption{(color online).
Comparison of [(a) and (b)] $v_2\{\Psi_2\}$ vs $N_{\rm part}$ 
and [(c) and (d)] $v_3\{\Psi_3\}$ vs $N_{\rm part}$ 
measurements and theoretical predictions (see text): 
``MC-KLN + $4\pi\eta/s = 2$" 
and ``Glauber + $4\pi\eta/s = 1$ (1)"~\protect\cite{Alver:2010dn}; 
``Glauber + $4\pi\eta/s = 1$ (2)"~\protect\cite{Schenke:2010rr}; 
and ``UrQMD"~\protect\cite{Petersen:2010cw}.  Shaded areas (magenta) 
around the data points indicate sizes of systematic uncertainties.
}
\label{Fig3}
\end{figure}

Figure~\ref{Fig3} compares the centrality dependence of
$v_{2}\{\Psi_2\}$ and $v_{3}\{\Psi_3\}$ with several additional
calculations, demonstrating both the new constraints the data provide
and also the robustness of hydrodynamics to the details of different
model assumptions for medium evolution.  Alver et
al.~\cite{Alver:2010dn} use relativistic viscous hydrodynamics in 2+1
dimensions. Fluctuations are introduced for two different initial
conditions. For Glauber initial conditions, the energy density
distribution in the transverse plane is proportional to a
superposition of struck nucleon and binary-collision densities; in
MC-KLN initial conditions the energy density profile is further
controlled by the dependence of the gluon saturation momentum on the
transverse position~\cite{Lappi:2006xc,Drescher:2007ax}. 
The Glauber-MC and MC-KLN initial state models are paired with the 
values $4\pi\eta/s = 1$ and 2, respectively, to reproduce the 
measured $v_{2}\{\Psi_2\}$~\cite{Song:2010mg}. The viscosity 
difference compensates for the $\sim$ 20\% difference between the initial $\varepsilon_2$ 
values associated with each model. The two models have similar $\varepsilon_3$, and 
thus the larger viscosity needed with MC-KLN to match $v_2$, 
leads to a much lower $v_3$ than obtained with Glauber-MC.
Consequently, our measurement of $v_{3}\{\Psi_3\}$ helps to
disentangle viscosity and initial conditions.  The efficacy of these
2+1 hydrodynamic results for Glauber initial conditions are confirmed
by further calculations with different model assumptions. Petersen et
al.~\cite{Petersen:2010cw} determine a Glauber initial state
event-by-event, translating through pre-equilibrium with the UrQMD
transport model~\cite{Bass:1998ca,Bleicher:1999xi}, then evolving
the medium with ideal QGP hydrodynamics ($\eta/s$ = 0), and finally
switching to a hadronic cascade (which has an effective viscosity) as
regions become dilute. B. Schenke {\it et al.}~\cite{Schenke:2010rr} use
event-by-event Glauber initial conditions, evolved with relativistic viscous
3+1 dimensional hydrodynamics with $4\pi\eta/s$ = 1.

%%%%%%%%%%%%%%%%%%%%%%%%%%%%%%%%%%%%%%%%%%%%%% Fig_3

All of these models are compared with $v_{2}\{\Psi_2\}$, and
$v_{3}\{\Psi_3\}$ data as a function of $N_{\rm part}$ in two $p_T$ bins.
All calculations describe $v_{2}\{\Psi_2\}$ well at $p_T$ = 0.75
GeV/$c$.  Deviations from hydrodynamics should be expected in
peripheral collisions, where nonequilibrium effects may be large.
At higher $p_T$, differences between the calculations become more
apparent.  All models still agree with $v_{2}\{\Psi_2\}$, including
MC-KLN initial conditions.  However, the lower panels of
Fig.~\ref{Fig3} show the constraining power of $v_{3}\{\Psi_3\}$
and that the calculated results from viscous
hydrodynamics, with MC-KLN initial conditions and $4\pi\eta/s
= 2$, lie significantly below the data. This is more apparent in the
higher $p_T$ bin, even in the most central collisions. Therefore, our
comparisons suggest that the combination of the current implementation
of the MC-KLN initial conditions
in concert with $4\pi\eta/s = 2$ is disfavored by our new
$v_{3}\{\Psi_3\}$ measurements.  This may suggest that the MC-KLN 
implementation or its application needs to be reevaluated 
(see~\cite{Albacete:2011fw}), but it does not neccessarily
imply that a CGC initial state is disfavored.

The results from the
hydrodynamical calculations which employ Glauber initial condition
fluctuations and $4\pi\eta/s = 1$ show relatively good
agreement with the $v_{2,3}\{\Psi_{2,3}\}$ data.  The exact
statistical significance of these constraints should be determined
through a global fit procedure, including a quantitative accounting of the
breakdown of hydrodynamics in peripheral collisions, as well as of the
systematics associated with the averaging of eccentricity
fluctuations within the models. From our data
it is already clear that the higher order moment $v_3$ should provide
an important avenue for constraining different physical properties of the QGP.

In summary, we have presented participant event plane $\Psi_n$
correlations and differential measurements of $v_n\{\Psi_n\}$ for
$n=2,3,4$ for charged hadrons using the generalized event-plane
method.  The higher order harmonic moments $v_3\{\Psi_3\}$ and 
$v_4\{\Psi_4\}$ and the nonzero correlations between higher-order 
event planes across a large rapidity gap of $\Delta\eta'{\gtrsim}7$,  
indicate that the initial state has transverse geometry 
fluctuations.  These fluctuations affect the generalized 
eccentricities, which are subsequently propagated in the 
hydrodynamic evolution of the plasma.
The evidence, includes (1) a lack of correlation between the
measured event planes of order $n=2$ and $3$ as predicted by Glauber
modeling, assuming correlations of the event planes with the
generalized eccentricity, (2) proper description of the shapes of
the $p_T$ dependence in the low $p_T$ region by hydrodynamic
calculations, and (3) agreement with several different initial state
+ hydrodynamic models across centralities for order $v_n\{\Psi_n\}$.
The combined results for $v_{2,3}\{\Psi_{2,3}\}$, together with initial 
hydrodynamic-model calculations now suggest that one of the important 
factors contributing to a large uncertainty in the extracted value of 
$4\pi\eta/s$ can be significantly reduced.  For the limited set 
of models considered, $4\pi\eta/s \simeq 1$ is favored.

%%%%%%%%%%%%%%%%%%%%%%%%%%  Acknowledgements  %%%%%%%%%%%%%%%%%%%%%%%%%%

%\section{Acknowledgements}   % Run-7 long from for PRC, PLB, etc.

We thank the staff of the Collider-Accelerator and Physics
Departments at Brookhaven National Laboratory and the staff of
the other PHENIX participating institutions for their vital
contributions.  We acknowledge support from the 
Office of Nuclear Physics in the
Office of Science of the Department of Energy, the
National Science Foundation, Abilene Christian University
Research Council, Research Foundation of SUNY, and Dean of the
College of Arts and Sciences, Vanderbilt University (U.S.A),
Ministry of Education, Culture, Sports, Science, and Technology
and the Japan Society for the Promotion of Science (Japan),
Conselho Nacional de Desenvolvimento Cient\'{\i}fico e
Tecnol{\'o}gico and Funda\c c{\~a}o de Amparo {\`a} Pesquisa do
Estado de S{\~a}o Paulo (Brazil),
Natural Science Foundation of China (P.~R.~China),
Ministry of Education, Youth and Sports (Czech Republic),
Centre National de la Recherche Scientifique, Commissariat
{\`a} l'{\'E}nergie Atomique, and Institut National de Physique
Nucl{\'e}aire et de Physique des Particules (France),
Ministry of Industry, Science and Tekhnologies,
Bundesministerium f\"ur Bildung und Forschung, Deutscher
Akademischer Austausch Dienst, and Alexander von Humboldt Stiftung (Germany),
Hungarian National Science Fund, OTKA (Hungary), 
Department of Atomic Energy and Department of Science and Technology (India), 
Israel Science Foundation (Israel), 
National Research Foundation and WCU program of the 
Ministry Education Science and Technology (Korea),
Ministry of Education and Science, Russian Academy of Sciences,
Federal Agency of Atomic Energy (Russia),
VR and the Wallenberg Foundation (Sweden), 
the U.S. Civilian Research and Development Foundation for the
Independent States of the Former Soviet Union, 
the US-Hungarian Fulbright Foundation for Educational Exchange,
and the US-Israel Binational Science Foundation.

%%%%%%%%%%%%%%%%%%%%%%%%%%%%%  References  %%%%%%%%%%%%%%%%%%%%%%%%%%%%%%

%%\bibliography{ppg132x2}

\end{document}